\documentclass{aa}

\usepackage{graphicx}
\usepackage{txfonts}
\usepackage{graphicx}   % Including figure files
\usepackage{amsmath}
\usepackage[nice]{nicefrac}
\usepackage{booktabs}
\usepackage{tabu}
\usepackage{siunitx}
\usepackage{makecell}
\usepackage{multirow}
\usepackage{float}
\usepackage{color}
\usepackage{listings}
\usepackage[colorlinks=true,     linkcolor=blue, citecolor=blue, filecolor=blue, urlcolor=blue]{hyperref}
\newcommand{\lime}{\textsc{LiMe}}
\let\oldAA\AA
\renewcommand{\AA}{\text{\normalfont\oldAA}}
\nolinenumbers

\begin{document}

   \title{\lime{}: A \emph{\textsc{Li}}ne \emph{\textsc{Me}}asuring library for large and complex spectroscopic data sets}

   \subtitle{I. Implementation of a virtual observatory for JWST spectra}

   \author{V. Fern\'andez \inst{1}\fnmsep\inst{2}, R. Amor\'in \inst{3}\fnmsep\inst{1}, V. Firpo \inst{4} and C. Morisset \inst{5}\fnmsep\inst{6}}
    
   \authorrunning{V.F., R.A., C.M., V.F}
    
   \institute{Departamento de Astronom\'ia, Universidad de La Serena, Av. Juan Cisternas 1200 Norte, La Serena, Chile\\
              \email{vital.fernandez@userena.cl}
        \and
             Michigan Institute for Data Science, Unversity of Michgigan, 500 Church Street, Ann Arbor, MI 48109, US
        \and
             ARAID Foundation, Centro de Estudios de F\'isica del Cosmos de Arag\'on (CEFCA), Unidad Asociada al CSIC, Plaza San Juan 1E, 44001 Teruel, Spain
        \and
             Gemini Observatory/NSF’s NOIRLab, Casilla 603, La Serena, Chile
        \and
             Universidad Nacional Autonoma de M\'exico, Instituto de Astronom\'ia (IA), Apdo. postal 106, C.P. 22800 Ensenada, Baja California, M\'exico
        \and
             Instituto de Ciencias F\'isicas, Universidad Nacional Autonoma de M\'exico, Av. Universidad s/n, 62210 Cuernavaca, Mor., M\'exico 
        }

   \date{Received September 15, 1996; accepted March 16, 1997}

% \abstract{}{}{}{}{} 
% 5 {} token are mandatory
 
  \abstract
   {The upcoming generation of telescopes, instruments, and surveys is poised to usher in an unprecedented "Big Data" era in the field of astronomy. Within this context, even seemingly modest tasks such as spectral line analyses could become increasingly challenging for astronomers.}
   {In this paper, we announce the release of \lime{}. This package is tailored for multidisciplinary observations with long-slit and integral field spectroscopy (IFS) support. \lime{} functions encompass the reading of  observational files, detecting lines, conditioned line fitting, and the plotting and storage of results. Most importantly, these measurements are structured to support the subsequent chemical and kinematic analyses.}
   {To reduce the coding effort required from users, we introduced a notation system for atomic transitions that is accessible to  humans and machine-readable. Along with this system, we present an extensive database of line bands, spanning from the ultraviolet to the infrared wavelength range. Additionally, we propose a model designed to train machine learning algorithms in line detection. \lime{} features a comprehensive online documentation, which details the command attributes and includes several tutorials. These tutorials range from measuring a single line to analyzing an entire IFS data cube.}
   {This library functions and measurements are showcased in an online virtual observatory. The data in this interactive website come from the JWST NIRSpec observations of the CEERs survey. In this regard, \lime{} offers improvements related to the dissemination and accessibility of astronomical spectra.}
   {}


   \maketitle

%-------------------------------------------------------------------
\nolinenumbers

\section{Introduction}

The term "spectrum" originates from the Latin verb "specere," meaning "to look." Over time, this meaning evolved to refer to a sudden image or apparition. \cite{newton_opticks_1704} coined this term to describe the colorful dispersion of light, when he demonstrated that it was not an intrinsic property of the prism but a manifestation of light itself. This image, however, was not as continuous as it seemed. During the measurement of the refractive power of various substances, via the implementation of the "camera lucida", \cite{wollaston_xii_1802} encountered a "well defined line, free from colour." The author argued that these were boundaries in the spectrum colours. However, a decade later, Fraunhofer's development of the modern spectroscope has enabled observations of almost 600 lines. In another remarkable contribution, \cite{fraunhofer_bestimmung_1815} compared the solar spectrum against the one from Sirius. The author noted that "these stars, in regard to the stripes, seem to differ among themselves." The implementation of spectroscopes and spectrographs in telescopes had critical repercussions in the astronomical field and beyond. In this regard, we refer to the research of William and Margaret Huggins. \cite{huggins_xiii_1864} were the first to use a "spectrum apparatus" to observe several nebulae. The authors described how at first they thought there was a "derangement of the instrument" since "no spectrum was seen but only a short line of light perpendicular to the dispersion direction". The authors correctly linked the nebula's "faintest" line to the hydrogen dark Fraunhofer F line ($H\beta$). The astronomers argued that these objects were not "aggregations of suns" but "enormous masses of luminous gas" since the emission line spectrum is characteristic of this medium. In "The Origin of the Nebulium Spectrum", \cite{bowen_origin_1927} linked these unknown lines to "familiar atoms" under "unfamiliar conditions". This condition is the extreme low density, which makes possible the natural de-excitation of collisionally excited electrons. Particle physics also explains the observed profiles of these lines with the arrival of high resolution spectrographs. Natural and pressure broadening are responsible for Lorentzian and Cauchy profiles, while thermal or Doppler broadening is responsible for Gaussian profiles. In particle transitions, where both mechanisms have similar weights, we observe the convolution of both profiles: a Voigt line \citep[see][]{armstrong_spectrum_1967}. 
Since the intensity and shape of these features depend on the gas chemistry and kinematics respectively, much of what humanity will ever know about the  composition and motion of  the Universe starts with the analysis of these lines. 

In the past two centuries, scientific developments in optical, electrical, and computational disciplines have translated into remarkable achievements in spectrograph design. In observational astrophysics, integral field spectroscopy (IFS) and massively- multiplexed spectroscopy, arguably provide the most valuable datasets: a uniform spectra sample covering a large field. Moreover, novel spectrograph designs can support both a large wavelength range and high resolution. Upcoming spectroscopic surveys are expected to produce more data per day than the complete historical release in past surveys \citep[see][]{zhang_astronomy_2015}, opening up a new era in our understanding of astronomical phenomena. On the negative side, this will involve the measurement of many lines. Additionally, a single Gaussian fitting \citep[see][]{mezger_galactic_1967,smith_internal_1970,hippelein_turbulent_1986,chu_internal_1994} is no longer sufficient as the new instruments reveal more kinematic components \citep[see][]{hagele_high-resolution_2012,bosch_integral_2019,hogarth_chemodynamics_2020}. This means that not only do we have more spectra to analyze, but their analyses must be tailored to a range of transitions.

To support the community, the authors have developed the \textbf{Li}ne \textbf{Me}asuring package: \lime{}. This library design is aimed at facilitating multi-disciplinary research, with support for long-slit and echelle spectra and IFS data cubes, line detection, emission and/or absorption profiles, integrated and profile fluxes, multi-component and multi-boundary fittings, pixel flux-uncertainty, and pixel masks. The package includes several tools to plot and interact with the input spectra and the output measurements. Furthermore, it features a simple installation procedure, comprehensive documentation, and a modern testing framework.

Measurements made using \lime{} have been utilized in the chemical and kinematic analysis of the extreme emission line galaxy CGCG007-025 by \cite{fernandez_new_2022}, \cite{delvalle-espinosa_spatially_2023}, and \cite{amorin_ubiquitous_2024} as well as in observations from the James Webb Space Telescope (JWST) by the Cosmic Evolution Early Release Science survey (CEERS) \citep[see][]{finkelstein_long_2022}. These initial applications have had a significant impact on the development of the \lime{} methodology. We seize this opportunity to introduce the line naming which effectively describes transition properties and a line bands database, which covers from the ultraviolet to the infrared. Lastly, we present a theoretical model designed to train machine learning algorithms for the automatic identification of lines. All these features are illustrated with an interactive website to visualize the spectra from the CEERs field with over 600 galaxies in the $0.29 \leq z \leq 9.62$ redshift range.

%--------------------------------------------------------------------
\section{Technical description}

\begin{figure*}
\centering
\includegraphics[width=1\textwidth]{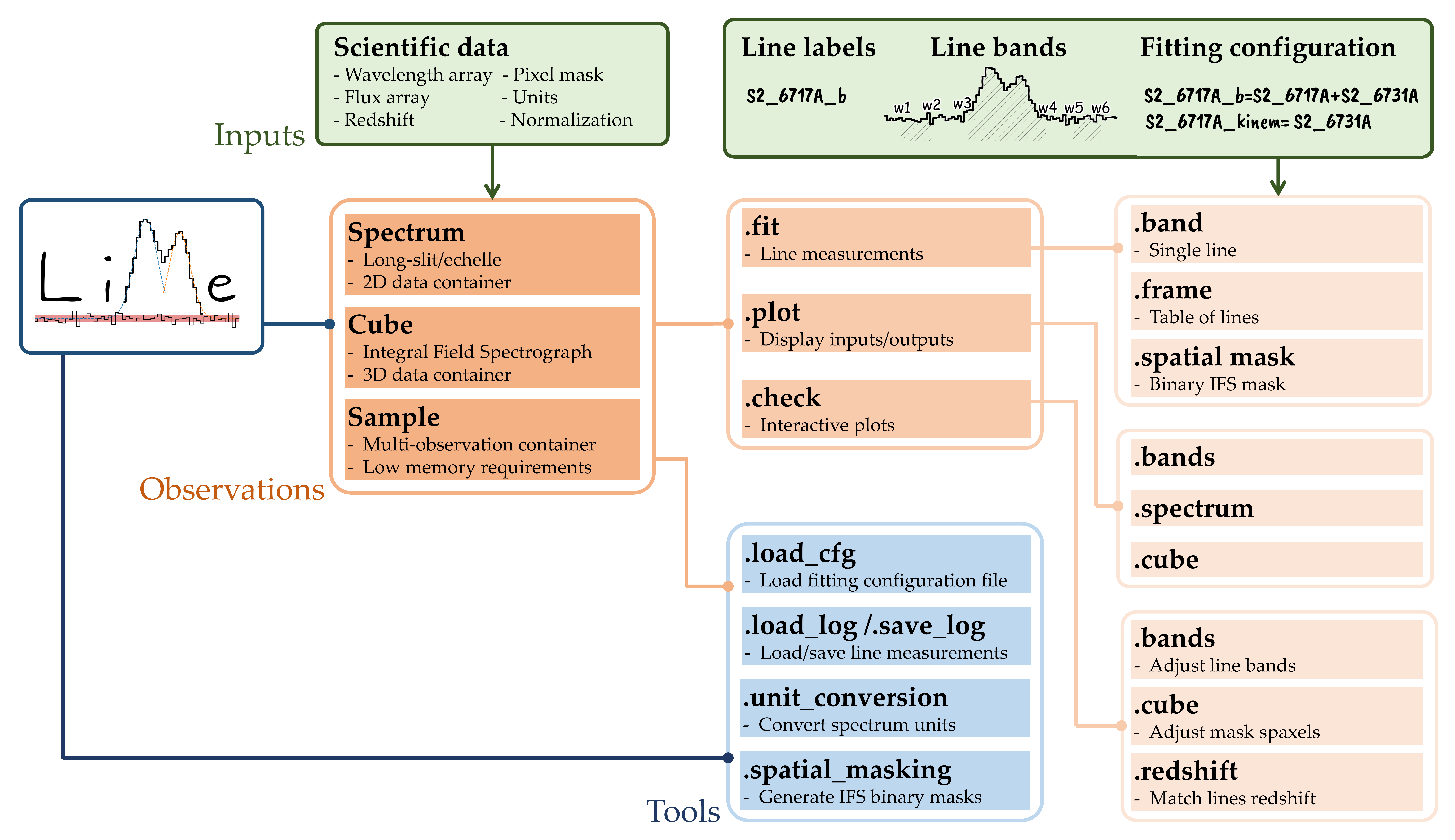}
\caption{\label{fig:workflow} \lime{} structure with some of the available functions. In the color version of the figure, the green background represents the library inputs, the orange cells cover the observations workflow, and the blue cell includes auxiliary functions}
\end{figure*}

\lime{} is developed in \textsc{Python} and is compatible with its latest version (v3.11). The numerical arrays are managed using \textsc{NumPy} \citep{harris_array_2020}, while minimization operations employ \textsc{LmFIT} syntax \citep{newville_lmfit_2014}. The plots use \textsc{Matplotlib} \citep{hunter_matplotlib_2007}, and measurements are stored as \textsc{Pandas} \emph{DataFrames} \citep{the_pandas_development_team_pandas-devpandas_2023}. This library also facilitates saving the measurements in \emph{.txt} and \emph{.csv} file formats. The \textsc{AstroPy} package \citep{collaboration_astropy_2022} is employed for handling \emph{.fits} files, as well as for managing the World Coordinate System (WCS) in IFS observation coordinates. Lastly, for \textsc{Python} versions below 3.11, users are required to install the \textsc{tomli} package\footnote{available at \href{https://pypi.org/project/tomli/}{https://pypi.org/project/tomli/}} to read \textsc{Toml} configuration files\footnote{see \href{https://toml.io/en/}{https://toml.io/en/}}.

In addition to these core dependencies, \lime{} supports \emph{.pdf}, \emph{.xlsx}, and \emph{.asdf} output files via the \textsc{PyLatex}, \textsc{openpyxl}, and \textsc{asdf} pacakges respectively. Additionally, when \textsc{mplcursors} is installed, clicking on Gaussian profile figure displays a pop-up with data regarding the line measurements.
\lime{} can be installed from its \href{https://pypi.org/project/lime-stable/}{\textsc{PyPi}} repository, using the following command:

\begin{verbatim}
pip install lime-stable.
\end{verbatim}

The \lime{} version is included in all the output measurement files, regardless of their type. Users can install a specific version or upgrade to the latest version using these commands:
\begin{verbatim}
pip install lime-stable==1.0.0
pip install lime-stable --upgrade
\end{verbatim}

\lime{} documentation is compiled at each update online \footnote{Documentation \href{https://lime-stable.readthedocs.io/en/latest/}{https://lime-stable.readthedocs.io/en/latest}}. The documentation's tutorial section ranges from fitting a single line to analyzing an IFS cube from the MANGA survey \citep[see][]{blanton_sloan_2017}. These tutorials, created from a series of notebooks, are available for download (along with the equivalent python scripts and scientific data) from the main author's \textsc{GitHub} page \footnote{GitHub \href{https://github.com/Vital-Fernandez/lime}{https://github.com/Vital-Fernandez/lime}}.

In order to guarantee consistency on the measurements, as new features are added, \lime{} includes a testing framework, which compiles automatically online\footnote{Tests compilation \href{https://app.codecov.io/gh/Vital-Fernandez/lime}{app.codecov.io/gh/Vital-Fernandez/lime}}.
Prospective users are encouraged to submit any comments, requests, or bug reports\footnote{Reporting issues \href{https://github.com/Vital-Fernandez/lime/issues}{GitHub in the issues section}}.

%--------------------------------------------------------------------

\subsection{Library structure}

\lime{} features a composite software design, utilizing instances of other classes to implement the target functionality. This approach is akin to that of \textsc{IRAF} \citep[see][]{tody_iraf_1986}: Functions are organized into multi-level packages, which users access to perform the corresponding task. The diagram in Fig.\ref{fig:workflow} outlines this workflow.

At the highest level, \lime{} provides of observational classes: spectrum, cube, and sample. The first two are essentially 2D and 3D data containers, respectively. The third class functions as a dictionary-like container for multiple spectrum or cube objects. Moreover, as illustrated in Fig.\ref{fig:workflow}, various tools can be invoked via the \lime{} import for tasks, such as loading and saving data. Many of these functions are also within the observations.

At an intermediate level, each observational class includes the \emph{.fit}, \emph{.plot}, and \emph{.check} objects. The first provides functions to launch the measurements from the observation data. The second organizes functions to plot the observations and/or measurements, while the \emph{.check} object facilitates interactive plots, allowing users to select or adjust data through mouse clicks or widgets. In these functions, users must specify an output file to store these user inputs.

Finally, at the lowest level, we find the functions that execute the measurements or plots. Beyond the aforementioned functionality, the main distinction between these commands lies in the extent of the data they handle. For instance, the \href{https://lime-stable.readthedocs.io/en/latest/introduction/api.html#lime.workflow.SpecTreatment.bands}{\emph{Spectrum.fit.bands}} and \href{https://lime-stable.readthedocs.io/en/latest/introduction/api.html#lime.workflow.SpecTreatment.frame}{\emph{Spectrum.fit.frame}} commands fit a single line and a list of lines in a spectrum, respectively. Conversely, the \href{https://lime-stable.readthedocs.io/en/latest/introduction/api.html#lime.workflow.CubeTreatment.spatial_mask}{\emph{Cube.fit.spatial\_mask}} command fits a list of lines within a spatial region of an IFS cube. A comprehensive description of the library functions and their attributes is available in the \href{https://lime-stable.readthedocs.io/en/latest/introduction/api.html}{API documentation}.

\subsubsection{ Declaration of observations} \label{sec:observations_creation}

At present, \lime{} can generate observation objects described above directly from \textsc{ISIS}, \textsc{OSIRIS}, \textsc{MEGARA}, \textsc{SDSS}, \textsc{MANGA}, \textsc{MUSE}, \textsc{and NIRSPEC} \emph{.fits} files. For other instruments users are required to use \textsc{Python} to load the observation data. The documentation provides \href{https://lime-stable.readthedocs.io/en/latest/inputs/n_inputs1_spectra.html}{several examples} on how to read \emph{.fits} files. Regardless of the observation type, most inputs are shared:

\textbf{Wavelength, flux, and flux uncertainty arrays}: These numerical arrays contain the spectrum's dispersion axis, the energy density axis, and the energy density standard deviation. The latter uncertainty array must match the units of the input flux, but it is optional. \lime{} assumes that the spectroscopic data is in the observed frame. At the creation of the observation variable, users can specify the minimum and maximum dispersion axis values to crop the spectra arrays.

\textbf{Redshift}: Much of \lime{}'s automation require the object's cosmological redshift. This includes the line detection functions and the bands calculation. However, the measurements are still performed in the observed frame.

\textbf{Units}: By default, \lime{} assumes \lstinline{unit_wave} = \lstinline{"Angstrom"} $(\AA)$ and \lstinline{unit_flux} = \lstinline{"FLAM"} $\left(\nicefrac{erg}{s \cdot cm^{2} \cdot \AA}\right)$ for the dispersion and flux energy density units, respectively. However, users can specify the observation units using the online notation \href{https://docs.astropy.org/en/stable/units/standard\_units.html}{\textsc{AstroPy} notation}. Additionally, users can convert the spectra units and normalization by calling the \href{https://lime-stable.readthedocs.io/en/latest/introduction/api.html#lime.unit_conversion}{.unit\_conversion} function. Finally, the units can include a scale factor. The output files with the measurements will include the input spectrum units. 

\textbf{Normalization}: Spectra using the centimeter-gram-second system (cgs) may display a flux density several orders of magnitude below unity. Most minimizing algorithms struggle to fit a theoretical profile at such a scale. \lime{} computes a normalization for the observation, if none is provided. Similarly, users should be mindful of the spectral dispersion axis units (such as $\mu m$), where the wavelength resolution may be several orders of magnitude below unity. This normalization is removed from the output measurements.

\textbf{Pixel mask}: When declaring the observation, users can specify a boolean array for pixels to be excluded from the analysis. This can be used to tag bad pixels (for example, non-numerical values) or non-physical entries (such as 0 or negative values). \lime{} ignores these pixels during measurements but displays them in plots as red crosses. It must be noted that \lime{} uses \href{https://numpy.org/doc/stable/reference/maskedarray.generic.html}{\textsc{NumPy} masked arrays}, in which masked entries have a \emph{True} boolean value.

In addition to the core attributes mentioned above, the observation classes have optional inputs that may contribute to the measurements workflow:

At the \href{https://lime-stable.readthedocs.io/en/latest/introduction/api.html#lime.Spectrum}{Spectrum} declaration, users can specify a constant instrument's Full Width Half Maximum (FWHM) and the electron temperature $(T_{e})$. The FWHM quantifies the instrument's capacity to distinguish between two contiguous wavelength values. These inputs are used to compute the instrumental ($\sigma_{inst}$) and thermal ($\sigma_{th}$) velocity dispersion corrections.

At the \href{https://lime-stable.readthedocs.io/en/latest/introduction/api.html#lime.Cube}{Cube} declaration, users can include a world coordinate system. \lime{} expects an \href{https://docs.astropy.org/en/stable/units/standard\_units.html}{\textsc{AstroPy} WCS} object for this geometric transformation. Typically, this variable can be constructed directly from the \emph{.fits} file header. The plots from the \emph{Cube} object will incorporate the WCS to display the data. Furthermore, for spatial masks and line measurement \emph{.fits} files, \lime{} will store this WCS on the output file header.

At the \href{https://lime-stable.readthedocs.io/en/latest/introduction/api.html#lime.Sample}{Sample} declaration, users must define a \emph{load\_function}. This is a \textsc{Python} function that specifies how to read an observation data file into a \lime{} \emph{Spectrum} or \emph{Cube} object. Unlike the previous two observation types, a \emph{Sample} is recommended for platforms with limited RAM, such as online platforms or budget laptops. \lime{} loads the scientific data only when the user requests an observation from the \emph{Sample} log. Additionally, the \emph{Sample} class is recommended for reviewing line measurements from multiple spectra. A comprehensive example is provided in the \href{https://lime-stable.readthedocs.io/en/latest/tutorials/n_tutorial7_Sample_creation.html}{$7^{th}$ tutorial}.

%--------------------------------------------------------------------

\section{Fitting inputs}

\begin{figure}
\centering
\includegraphics[width=1\columnwidth]{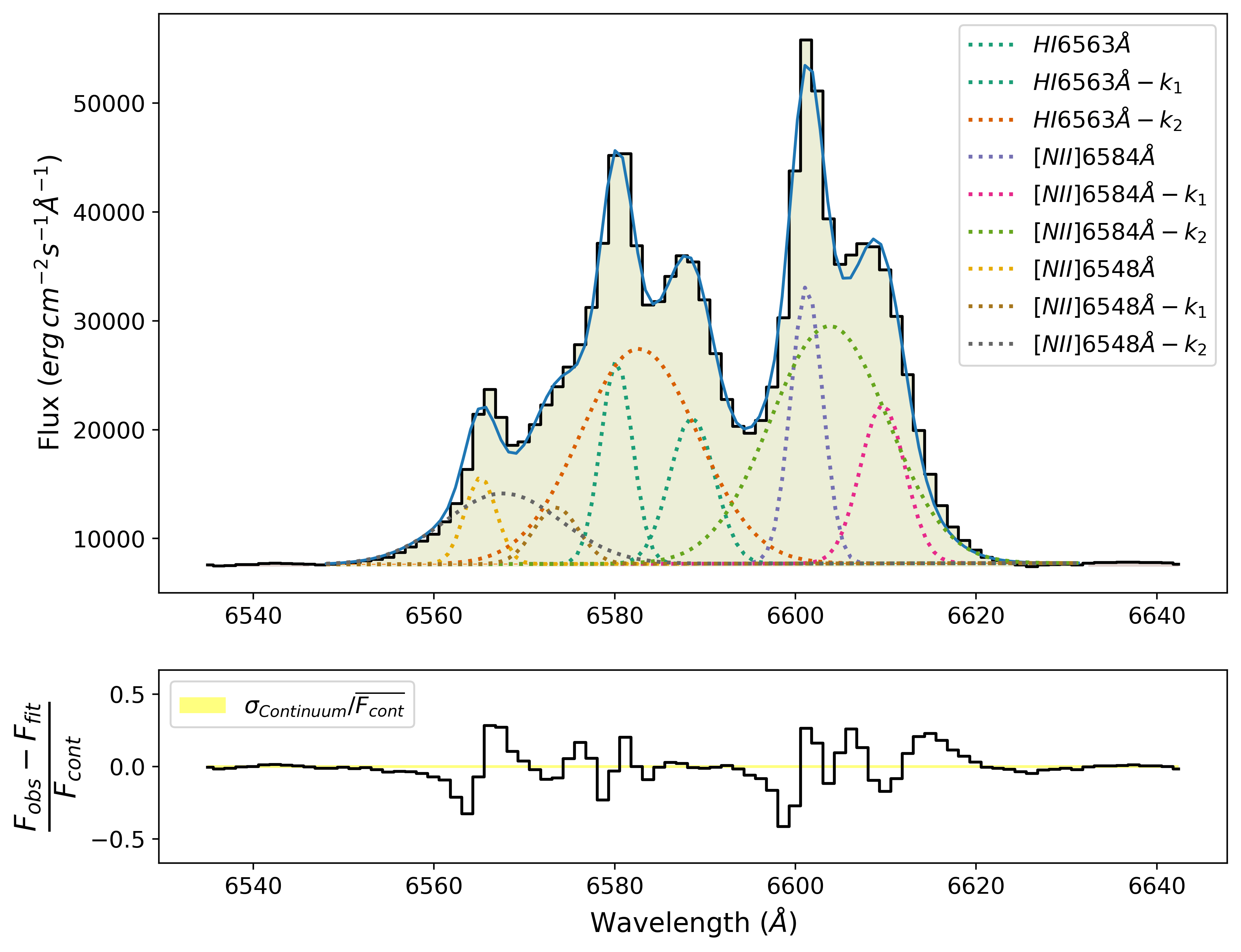}
\caption{\label{fig:Halpha_fitting} \lime{} fitting of the $H\alpha$ and $[NII]6548,6583\AA$ transitions in the MUSE observation of NGC1386 black hole region. In this fitting, we are exporting the kinematics of $[OIII]5007\AA$ components to the corresponding three transitions to facilitate the fitting.}
\end{figure}

As displayed in the top right corner of Fig. \ref{fig:workflow}, each line measurement in \lime{} requires three inputs: the line label, line bands, and fitting configuration. Unlike other packages, \lime{} has a strict definition for these inputs. In the recommended workflow, these inputs should be stored in external files: the line labels and bands in a tabulated file, and the fitting configuration in a \href{https://toml.io/en/}{\textsc{Toml}} text file. The rationale behind this design is evident in the $H\alpha-[NII]6548,6583\AA$ fitting shown in Fig. \ref{fig:Halpha_fitting}. This spectrum is from the MUSE (Multi Unit Spectroscopic Explorer) \citep[see][]{bacon_muse_2010} observation of NGC1386 supermassive black hole region. This data set was used in the workshop  publication titled \textit{Large-Volume Spectroscopic Analyses of AGN and Star Forming Galaxies in the Era of JWST} to compare results between different line measuring libraries \citep[see][]{stsci_large-volume_2022}. The reader can appreciate that there are three components for each of the three transitions. In the chemical and/or kinematic analysis of such a spectrum, it is crucial to remain consistent with the kinematic component label to avoid erroneous measurements. This complexity is even higher in IFS observations, where ionization and kinematic conditions vary from spaxel to spaxel. Explicit boundary conditions are therefore essential to ensure the minimizer assigns a consistent label to each profile parameter prior to the fitting. As a result, the transition notation must be highly informative to keep boundary conditions and output measurements organized. The subsequent sections describe how \lime{} addresses these challenges, while reducing the coding requirements for the user.

\subsection{Line labels} \label{sec:labels}

\begin{figure}
\centering
\includegraphics[width=1\columnwidth]{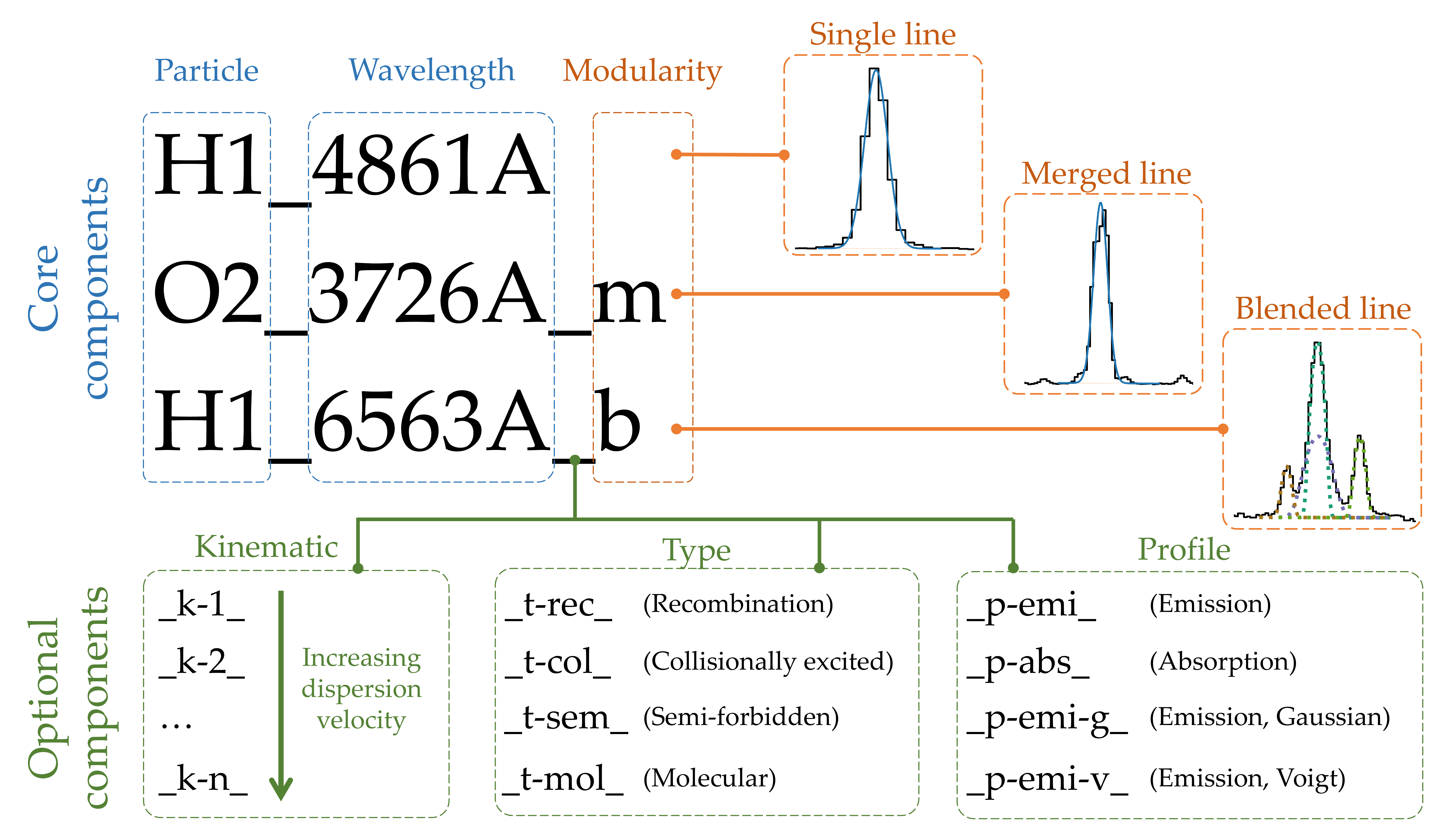}
\caption{\label{fig:line_notation} Line notation used in \lime{}. In the color version, the core components are displayed in blue, while the optional components are in green. The modularity suffix establishes the type of line components: \emph{Single} for just one transition in the profile. Multi-component transitions are labeled as \emph{Merged} if the components cannot be isolated, and  blended when  isolation is possible. }
\end{figure}

Although the notation for atomic transitions is well established \citep[see][and references therein]{osterbrock_astrophysics_1974}, it poses challenges for computational readability and processing. For starters, parsing roman numerals alongside alphabetic characters. Additionally, programming languages have limited alphanumeric characters\footnote{as indicated in the \href{https://unicode.org/reports/tr31/}{UAX \#31 unicode identifier and pattern syntax}}. This leaves many unit characters and mathematical operators without a symbolic representation. Finally, some characters like dots $(.)$, commas $(,)$, and square brackets $([])$ have strict definitions in programming scripts and configuration files. These limitations are well-known to researchers in software development. In private communications, the author engaged with developers of \textsc{Cloudy} \citep[see][]{chatzikos_2023_2023}, \textsc{HII-to-Chemistry} \citep[see][]{perez-montero_deriving_2014}, and \textsc{Fiasco} to discuss the line label formatting in their databases. The ultimate aim is to ensure that \lime{}'s notation is easily recognized by chemical and kinematic packages using its fluxes.

Fig. \ref{fig:line_notation} summarizes our transition notations. This is an expansion of the labeling used in \textsc{PyNeb} by \cite{luridiana_pyneb:_2015}. For example, the Balmer and Paschen alpha transitions are typified as \lstinline{H1_6563A} and \lstinline{H1_18750A} respectively. A line label is divided into components using underscores $(\_)$, while each component may be further split into items via dashes $(-)$. In \lime{}, there are three core components, which are compulsory and their order is fixed:

\begin{itemize}
    \item The first core component represents the \textbf{particle} responsible for the transition. By default \lime{} expects the particle chemical symbol followed by the ionization state in arabic numerals. This particle mass will be used to compute the thermal dispersion velocity in the output measurements. The user can add additional details to the transition by via dashses. For example: \lstinline{H1_18750A}, \lstinline{H1-PashchenAlpha_1875nm}, or \lstinline{H1-4-3_1875.0nm} are all processed similarly.

    \item The second item is the transition \textbf{wavelength}. This positive real number must be followed by the transition's wavelength or frequency units. These units must follow the \textsc{AstroPy} notation\footnote{see \href{https://docs.astropy.org/en/stable/units/standard\_units.html}{astropy units}}. The only exception is the "Angstrom" which in \href{https://docs.astropy.org/en/stable/units/standard\_units.html}{\textsc{AstroPy}} convention must be defined as "AA" or "Angstrom" but in this notation we can use "A". Importantly, this wavelength must be in the rest frame.

    \item The third core item is the line's \textbf{modularity}. This component, which must be last, even if there are optional components, indicates whether the label designates more than one particle transition. The component informs \lime{} on the profile fitting type. Fig. \ref{fig:line_notation} illustrates three possible scenarios: a "single line" is the default, where only one transition contributes to the line, and hence there isn't a modularity component; a "blended line" consists of two or more transitions, designated by the "\_b" suffix, where \lime{} fits one Gaussian profile per component in the fitting configuration; a "merged line," also consisting of two or more transitions and designated by the "\_m" suffix. In this case \lime{} fits a single Gaussian but the output measurements keep track of the components. This notation is useful in observations, where spectral resolution is insufficient to resolve individual components but posterior work, such as the fitting of photo-ionization models \citep[see][]{perez-montero_deriving_2014}, need this information. 

\end{itemize}

In contrast, the optional components in this notation have limited values. For example, the first item is a single string character that acts as a key for the component type. The order of these components can be arbitrary. Moreover, \lime{} will assign default values if they are not included in the transition. This design allows for future expansions, however, at present, the optional components are as follows:

\begin{itemize}
    \item The \textbf{kinematic} component: This first item is the letter "k", while the second one is the component cardinal number. In single and merged lines, \lime{} assumes the unique component is "0". Therefore, in blended lines, the user should name the second component \emph{k-1}, the third as \emph{k-2}, and so on. It's recommended to define these components from lower to higher dispersion velocity. However, users need to specify the boundary conditions in the fitting configuration to ensure this pattern.
    
    \item The \textbf{type} component: The first item is the "t" followed by a string specifying the particle transition type. Currently, \lime{} recognizes "t-rec" for recombination, "t-col" for collisionally excited, and "t-sem" for semi-forbidden lines. If this component is not provided, \lime{} will check the particle component and try to assign a recombination or collisionally excited tag automatically. This component is used to reconstruct the standard transition notation in \LaTeX.

    \item The \textbf{profile} component: The first item is the letter "p" followed by strings to declare the profile type. At present, \lime{} only fits Gaussian (g), Lorentz (l), pseudo-Voigt (pv) and exponential (e) in emission (emis) or absorption (abs). The default profile is a Gaussian curve in emission (\emph{\_p-g-emi}). The default profile can be changed with the arguments in the \emph{.fit.} commands.
    
\end{itemize}

% Further details and examples of this notation can be found in the \lime{} \href{\LEt{ footnote.***}https://lime-stable.readthedocs.io/en/latest/inputs/n_inputs2_line_labels.html}{inputs documentation}.

Further details and examples of this notation can be found in the \lime{} documentation\footnote{\href{https://lime-stable.readthedocs.io/en/latest/inputs/n_inputs2_line_labels.html}{Inputs documentation}.}.

\subsection{Line bands} \label{sec:line_bands}

As shown in Fig. \ref{fig:workflow} top right corner, the line bands constitute the second input in a line fitting. These is a 6-value array specifying the line spectral location and two bands of adjacent continua. It is essential that these array values are in the same units as the spectroscopic observation. Furthermore, it is recommended that the array is sorted. This design is based on the Lick system introduced by \cite{burstein_old_1984} and \cite{faber_old_1985}, used in the chemical analysis of absorption features in stellar spectra. These band-passes are limited to the $4000\AA \lesssim \lambda \lesssim 6500\AA$ wavelength range. Based on the authors' past observations, including JWST data described in Sect. \ref{sec:JWST_virtual}, we propose a set of bands covering lines from the ultraviolet to the infrared. The online supplementary material of this manuscript includes a text file with these bands. Additionally, users can call the \href{https://lime-stable.readthedocs.io/en/latest/introduction/api.html#lime.line\_bands}{lime.line\_bands} function to tailor the default database to specific species or transitions, spectrum range and units, and observation conditions (air/vacuum). The output table includes the line label as described in the previous section, with the requested formatting. Currently, this function does not interpolate the default bands to the spectrum resolution. As a workaround, users can manually adjust the bands, or remove bands for non-observed lines \href{https://lime-stable.readthedocs.io/en/latest/introduction/api.html#lime.line\_bands}{Spectrum.check.bands} function. This interactive plot also allows for the assignment of the line "single", "blended", and "merged" suffixes. In the recommended workflow, users declare these bands in a tabulated file where the rows are indexed by the line labels and the line band columns with the "w1", "w2", ... "w6" headers. Additionally, users can specify columns for the transition "wavelength" core component and the "kinematic", "type", "profile" optional components. This enables users to fine-tune the fitting configuration without lengthy transition labels. Moreover, a "latex\_label" column can be provided to override the default standard transition notation.

\subsection{Spatial masking} 

In the case of IFS observations, the user is encouraged to use spatial masks. These are matrices, where the \emph{True} entries correspond to spaxels where line measurements are to be performed. This design ensures that the analysis is limited to regions with scientific data. Additionally, masks can be used to adjust the fitting configuration for different objects or physical conditions within an IFS cube. \lime{} includes the \href{https://lime-stable.readthedocs.io/en/latest/introduction/api.html#lime.workflow.CubeTreatment.spatial\_mask}{Cube.check.spatial\_mask} function to generate spatial masks based on the signal-to-noise ratio (S/N) of the line bands. Currently, there are three options provided in this command:

\textbf{Band flux} (flux): This option sums up the input band flux across the spatial coordinates. The flux does not remove the continuum level in the case of an absorption or emission line. Users can specify flux percentiles to establish the number and extent of multiple masks.

\textbf{Continuum signal to noise} (SN\_cont): This option uses the input band to calculate the continuum S/N using the root mean square (RMS):
\begin{equation}
    \frac{S}{N}_{cont}=\left(\frac{1}{N}\sum_{i}^{N}\left(F_{mean}-F_{\lambda}\right)^{2}\right)^{\frac{1}{2}}
.\end{equation}
Here, $F_{\lambda}$ is the flux from each pixel in the band, $F_{mean}$ is the mean flux value from all the pixels, and $N$ is the number of pixels. Users can specify S/N percentiles to adjust the number and extent of the masks.

\textbf{Line S/N} (SN\_line): This option uses the line and adjacent continua bands to calculate the S/N of the input line using the definition by \cite{rola_estimation_1994}:
\begin{equation}
\frac{S}{N}_{line}\approx\frac{\sqrt{2\pi}}{6}\frac{A_{line}}{\sigma_{cont}}\sqrt{N} \label{eq:SN_rola}
.\end{equation}
Here, $A_{line}$ is the amplitude of the line peak, and $\sigma_{cont}$ is the sigma flux from the adjacent continua bands. For mask computation, the first term is calculated as the maximum pixel flux in the line band minus the mean continua flux. The second term is derived from the adjacent continua flux standard deviation. Users can specify S/N percentiles to adjust the number and extent of the masks.        

More details regarding the masking generation process can be found in the \href{https://lime-stable.readthedocs.io/en/latest/tutorials/n_tutorial4_IFU_masking.html}{$4^{th}$ \lime{} tutorial}.

\subsection{Fit configuration} \label{sec:fit_conf_levels}

\begin{figure}
\centering
\includegraphics[width=1\columnwidth]{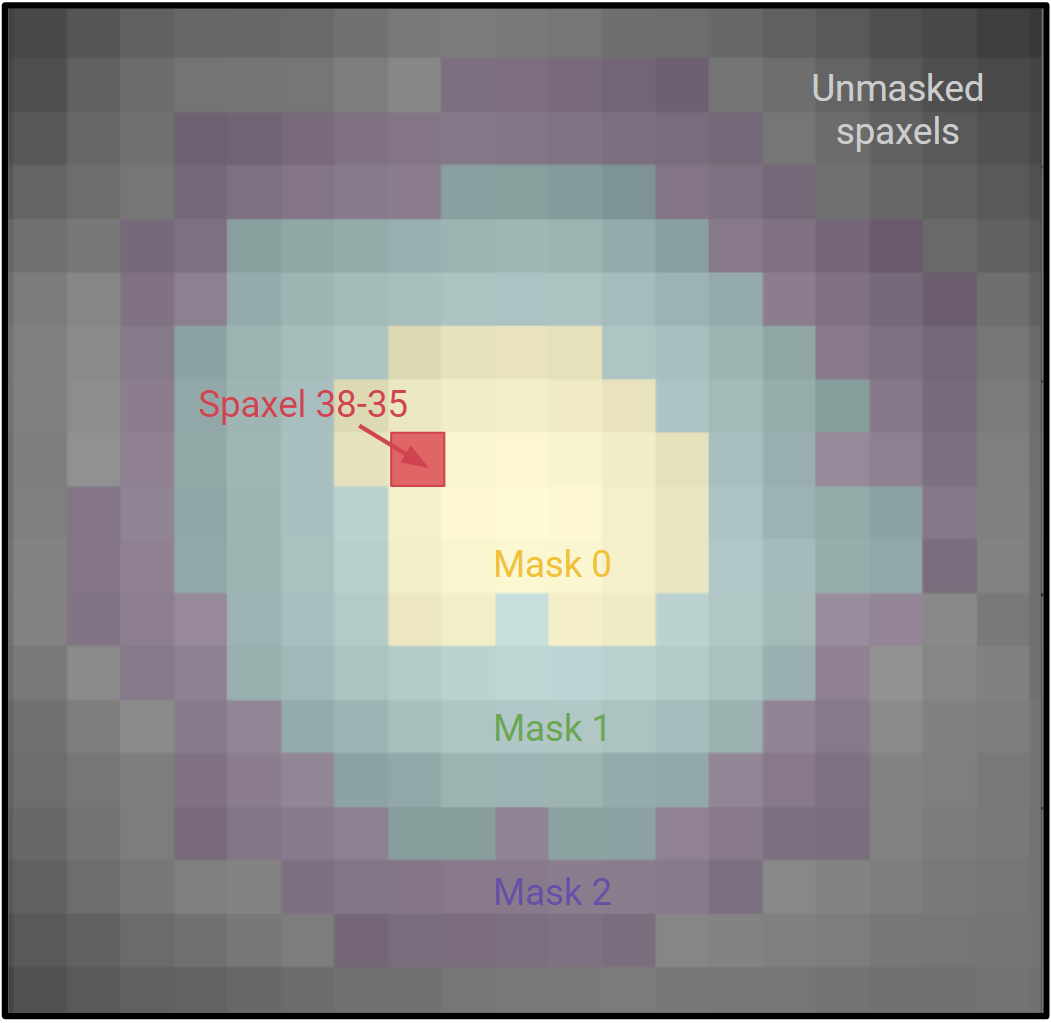}
\includegraphics[width=1\columnwidth]{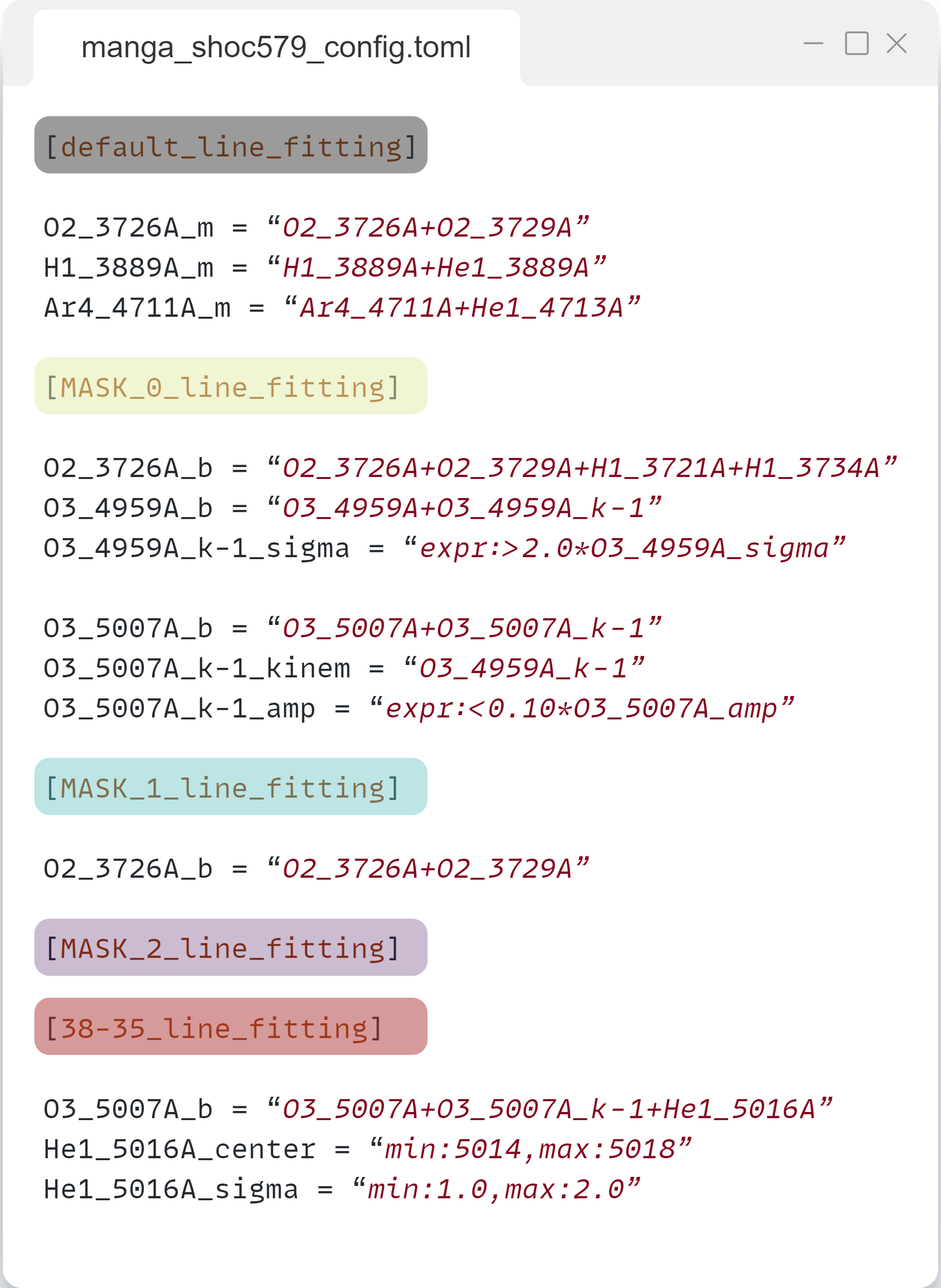}
\caption{\label{fig:fit_conf}  Spatial mask SHOC579 Manga observation (top). Part of the \lime{} configuration file for the IFU fitting (bottom).}
\end{figure}

Given the previous inputs, \lime{} has sufficient information to start measuring lines. In this scenario, however, all lines will be fitted with a single Gaussian. Indeed, even if the lines are correctly labeled with the merged ("\_m") and blended ("\_b") suffixes, they will still be treated as single transitions unless the components are explicitly specified. This is done via the fitting configuration files.

Fig. \ref{fig:fit_conf} illustrates an example of a fitting configuration file for analyzing an IFS observation. The configuration file follows a \href{https://toml.io/en/}{\textsc{Toml}} structure: The parameters are defined as key-value pairs, with different sections indicated by titles marked with square brackets. \lime{} just reads sections with the \emph{"\_line\_fitting"} suffix. This approach enables a multi-level configuration to adjust the line configuration while minimizing the user inputs. In the case of the \href{https://lime-stable.readthedocs.io/en/latest/introduction/api.html#lime.workflow.CubeTreatment.spatial_mask}{\emph{Cube.fit.spatial\_mask}}, the user has up to three levels to personalize the fittings:

\textbf{Default configuration}: These settings are applied to all the spaxels. As shown in Fig. \ref{fig:fit_conf}, the \emph{"default\_line\_fitting"} section includes the components for the merged and blended lines. The individual components must follow the notation described in Sect. \ref{sec:labels}, joined by (+) symbols.

\textbf{Mask configuration}: In Fig. \ref{fig:fit_conf}, there are three additional sections, one for each mask. In a star-forming galaxy, such as SHOC579 \citep[see][]{perez-montero_are_2013}, the localized radiation at the central stellar produces a rich emission spectra at the core. This emission gradually decreases towards the outskirts of the galaxy. Consequently, it is recommended to leverage \lime{}'s spatial treatment to adjust the analysis: moving from more lines with more complex profiles towards regions  with only single and merged lines as the S/N decreases. For spaxels within each mask, this information updates the fitting configuration from the default line fitting, inclusively. For example, for spaxels in the \emph{Mask\_2}, whose section \emph{[MASK\_2\_line\_fitting]} does not contain additional information, the \emph{[default\_line\_fitting]} would be used. In contrast, for \emph{Mask\_0} spaxels, the data in \emph{[MASK\_0\_line\_fitting]} would overwrite existing entries in the default configuration, while appending the new ones.

\textbf{Spaxel configuration}: In cases where certain spaxels need their own fitting configuration, the user can add a section titled with the spaxel array coordinates (Y-X, in \textsc{NumPy} array index). In Fig. \ref{fig:fit_conf}, the 38-35 spaxel inherits the \emph{default}, \emph{Mask\_0}, and \emph{38-35} line fitting configuration. Here, higher levels overwrite values from the lower levels. In this case, the \emph{O3\_5007A\_b} deblending at this spaxel will include the \emph{He1\_5016A} line as the spaxel blended components are updated from the \emph{Mask\_0} configuration.

The line fitting using the \href{https://lime-stable.readthedocs.io/en/latest/introduction/api.html#lime.workflow.SpecTreatment.frame}{\emph{Spectrum.fit.frame}} command in \lime{} offers two levels of configuration: global and local. This approach is advantageous for handling multiple spectra: a default global configuration can be applied to the entire sample, with a local configuration for specific objects.

At the present time, \lime{} can fit Gaussian, Lorentz, pseudo-Voigt, and exponential profiles. Gaussian is the default profile, if none is specified via the label profile suffix (see Fig. \ref{fig:line_notation}). As commented before, we use the \textsc{LmFit} by \cite{newville_lmfit_2014} to perform the fittings. However, we do not use the default models but our own expressions, which include the theoretical area and the FWHM for these profiles. For example, the Gaussian one is defined as:
\begin{equation}
F_{\lambda}=\sum_{i}A_{i}e^{-\frac{\left(\lambda-\mu_{i}\right)^{2}}{2\sigma_{i}^{2}}}+\left(m \lambda+n\right)
.\end{equation}
On the left-hand side, $F_{\lambda}$ represents the line flux. On the right-hand side, there are two components: the Gaussian profile flux, where $A_i$ is the height of the Gaussian profile above the continuum level, $\mu_i$ is the center of the Gaussian profile, and $\sigma_i$ is the standard deviation of the profile. The subscript $i$ denotes the transition. In the fitting configuration, these parameters are labeled with the name of the line, plus the suffixes \lstinline{_amp}, \lstinline{_center}, and \lstinline{_sigma} respectively. The second component is the continuum, modeled as a linear profile, where $m$ is the gradient and $n$ the intercept. In the configuration logs, these parameters are named after the blended line with the \lstinline{_slope} and \lstinline{_intercept} suffixes, respectively. Regardless of the number of components, there is only one continuum level, with the continuum linear parameters fixed by default from values derived from the adjacent bands continuum analysis carried out before the profile fitting. The online documentation \footnote{Profile fitting \href{https://lime-stable.readthedocs.io/en/latest/inputs/n_inputs4_fit_configuration.html}{documentation}} provides a comprehensive description of the options available to adjust the profile fitting. However, the following paragraphs summarize the key features:

    \textbf{Parameter boundaries:} In Fig. \ref{fig:fit_conf}, we see parameters constrained with \lstinline{min} and \lstinline{max} boundaries. This limits the parameter value range during the fitting. Additionally, the user can specify an initial \lstinline{value} for the fitting. If the attribute \lstinline{vary=False} the parameter's initial value remains fixed during the profile fitting. These attributes correspond to \href{https://lmfit.github.io/lmfit-py/parameters.html}{LmFIT parameters} declaration. For the Gaussian line center, constraints should be introduced in the rest-frame, and \lime{} will transform them to the observed-frame.
    
    \textbf{Intra-parameter boundaries:} For blended lines, the user can include the \lstinline{expr} attribute in parameter constraints. This attribute takes a mathematical string expression involving the parameter values. This relation will be enforced during the fitting. In Fig. \ref{fig:fit_conf}, we have an example where we use an expression to keep the $[OIII]4959,5007\AA$ amplitude fixed to their theoretical emissivity ratio. The \lstinline{expr} attribute is limited to blended lines. This ensures that  conflicts in samples, where the same line can be a found in blended group, can be avoided. This is a small trade-off for single lines since the \lstinline{expr} attribute is unlikely to be necessary.

    \textbf{Inequality boundaries:} The user can include inequality symbols in the \lstinline{expr} value to define boundaries, as seen in Fig. \ref{fig:fit_conf} for \lstinline{O3_4959A_k-1_sigma}. This inequality ensures that the kinematic component \lstinline{_k-1} maintains a larger magnitude than the narrower \lstinline{O3_4959A}. This enhancement is built over the \textsc{LmFIT} syntax by \lime{} and currently only supports a multiplicative factor.
 
    \textbf{Pixel mask}: In addition to global pixel masks (Sect. \ref{sec:observations_creation}), users can specify a pixel mask for a specific fitting. These keys are specified with the line name plus \lstinline{_mask}. The user can specify a single wavelength value or a two-limits interval with dash-separated floats, and combine multiple values or intervals with underscores.

    \textbf{Export line kinematics}: Including a line with the \lstinline{_kinem} suffix allows the user to match the kinematics of another line. This transformation occurs in the velocity plane, considering the Doppler effect. The child line kinematics are defined as:
    \begin{eqnarray}
    \mu_{child} & \left(\text{Å}\right)= & \mu_{parent}\cdot\frac{\lambda_{child}}{\lambda_{parent}},\\
    \sigma_{child} & \left(\text{Å}\right)= & \sigma_{parent}\cdot\frac{\lambda_{child}}{\lambda_{parent}},
    \end{eqnarray}
    The parent line can be from a previous measurement or in the current one. In the former case, the parent line must be measured in advance, and the Gaussian centroid and standard deviation remain fixed during fitting. In blended lines, parent and child line kinematics are tied during the fitting. This methodology was first introduced by \cite{bosch_integral_2019}.

%--------------------------------------------------------------------

\section{Line detection, measurement, and outputs }

In an ideal world, astronomers would have the time to individually check each spectrum before performing measurements. However, in the current age of Big Data, this is no longer feasible. As hinted in the previous section, one alternative is to group data into sub-samples of spectra or spatial regions sharing a common fitting configuration. Another critical aspect of automation is the detection of lines prior to their measurement. In this section, we present the peak or trough detection algorithm and line measurements.

\subsection{Emission and absorption detection} \label{sec:line_detec}

As outlined in the review by \cite{yang_comparison_2009}, the classical methodology to detect peaks on spectra involves three steps. The first is a smoothing operation, where the spectrum is convolved by a Gaussian filter to artificially remove noise from the continuum. The second step is normalization: the spectrum is divided by a polynomial fit of the continuum to produce a flattened continuum. Finally, the actual peak/trough detection involves passing the smoothed and normalized spectrum through a noise threshold. Flux points above/bellow these limits indicate an emission/absorption line. This is the very approach used by \cite{astropy-specutils_development_team_specutils_2019}. In \lime{} this procedure is managed by the following tasks:

    \href{https://lime-stable.readthedocs.io/en/latest/introduction/api.html#lime.workflow.SpecTreatment.continuum}{Spectrum.fit.continuum}: This command provides the iterative fitting of the continuum. The user declares two arrays of the same length: one with the polynomial order and the other, the intensity threshold factor. At each iteration, the code fits the continuum and masks pixels above/below the flux standard deviation times the intensity threshold factor. The aim of this loop is to gradually increase the polynomial order while decreasing the intensity threshold factor, thereby excluding emission/absorption line pixels for a better continuum fit. The user can also include a Gaussian filter size to smooth the spectrum. Upon completion, \emph{lime.Spectrum.continuum} and \emph{lime.Spectrum.cont\_std} attributes retain an array with the fitted continuum flux and the standard deviation from the observed minus fitted continua, respectively.

    \href{https://lime-stable.readthedocs.io/en/latest/introduction/api.html#lime.Spectrum.line\_detection}{Spectrum.line\_detection}: This command performs two operations. Initially, it normalizes the observation by dividing it by the previously calculated continuum. Then, it employs the \textsc{SciPy} task \href{https://docs.scipy.org/doc/scipy/reference/generated/scipy.signal.find_peaks.html}{find\_peaks}, using the pre-calculated continuum flux standard deviation as the intensity threshold times the input multiplicative factor. Afterwards, the resulting peaks  and troughs are cross-referenced with the user’s input lines database, which should adhere to the bands format described in section \ref{sec:line_bands}. Peaks/troughs within a line band receive a positive detection label. The output is a cropped version of the input bands log, containing only the detected lines.

The users can specify the attributes for these tasks within the configuration file. As in the case of the line fitting configuration described in Sect. \ref{sec:fit_conf_levels}, these attributes can be assigned in a multi-level priority that \lime{} will automatically assign for the corresponding spectrum and/or spatial mask.

\subsection{Output measurements} \label{sec:measurements}

\begin{figure*}[h]
\includegraphics[width=1\textwidth]{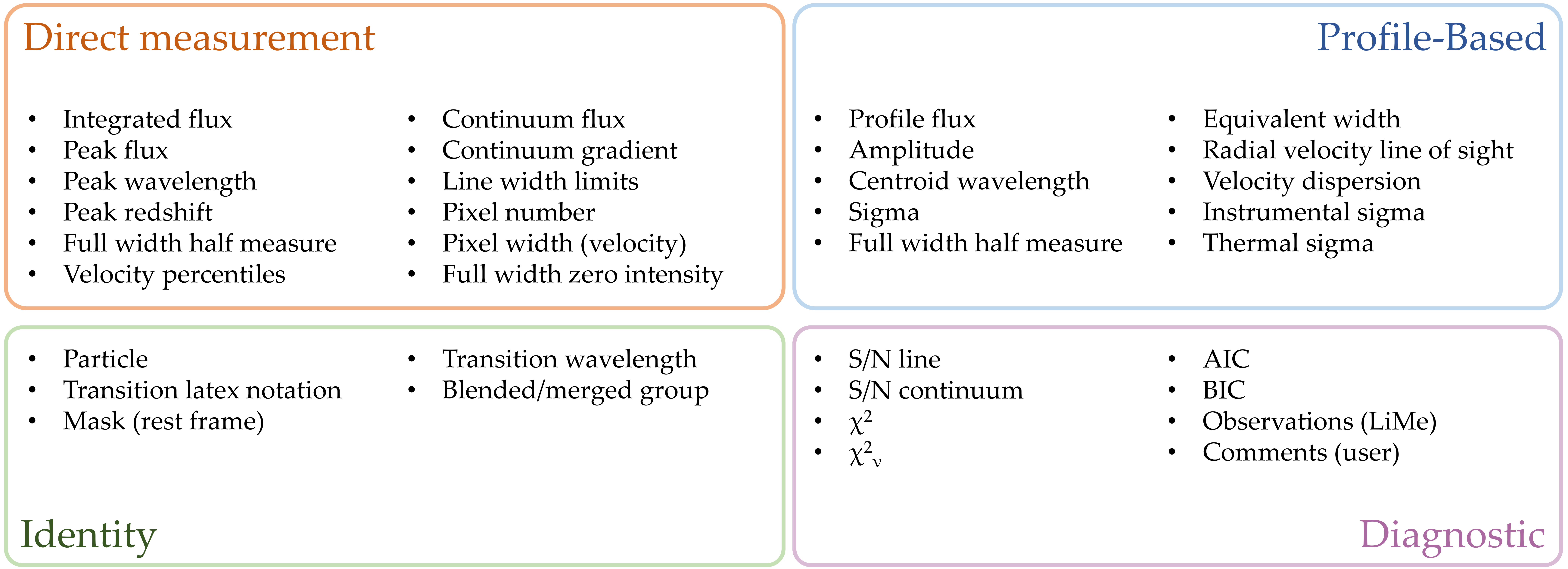}
\caption{\label{tab:measurements} \lime{} measurements division into several categories. Direct measurements are calculated from the integration of the line bands. Profile-based measurements depend on the theoretical profile assumption. The identity entries include information regarding the particle transition. Finally, the diagnostic measurements provide information regarding the quality of the fittings. A more detailed information can be found online \href{https://lime-stable.readthedocs.io/en/latest/outputs/outputs1_measurements.html}{ \textsc{LiMe} measurements documentation}}
\end{figure*} 

Fig. \ref{tab:measurements} provides a summary of the line measurements produced by \lime{} after each fitting. Furthermore, the online documentation\footnote{Online \href{https://lime-stable.readthedocs.io/en/latest/outputs/outputs1_measurements.html}{measurements description}} offers a detailed explanation of how these parameters are calculated. Therefore, this section will focus on the main characteristics of \lime{} measurements.

\begin{itemize}
    \item \lime{} yields two flux measurements for every line fitting. The first is an integration of the line bands pixels, via a bootstrap algorithm. In this process, every pixel receives a stochastic flux from a Gaussian distribution centered at zero, whose width comes from the pixel flux uncertainty. This is executed in a 1000-step loop. The output integrated flux and uncertainty are derived from the mean and standard deviation of these randoms arrays. In contrast, the profile flux is calculated from the theoretical relation. For a multi-Gaussian profile, we have, for example:
    \begin{equation}
    F_{i,g}=A_{i}\cdot\sqrt{2\pi}\cdot\sigma_{i} \label{eq:Gauss}
    .\end{equation}
    
    The Gaussian flux uncertainty depends on the algorithm chosen for the minimization\footnote{The minimization algorithm is specified in the fitting functions via the \lstinline{min\_method} attribute, using \href{https://lmfit.github.io/lmfit-py/fitting.html\#lmfit.minimizer.Minimizer.minimize}{\textsc{LmFIT} keywords}. The default option, least-squares minimization using the Trust Region reflective method, seems to provide the most stable results for complex boundary conditions.}. In most cases, this uncertainty is the standard error, indicating how much the parameter value must increase to raise the $\chi^2$ value by one unit. Both flux measurements are in the observed frame. 

    \item If the user provides the spectrum flux uncertainty as detailed in section \ref{sec:observations_creation}, \lime{} considers the individual pixel error in the flux calculations. If not, \lime{} calculates a uniform pixel flux uncertainty for each line, assuming a linear continuum from the adjacent bands' continua.
     
    \item Reference velocity: The radial velocity in the line of sight $(v_r)$ and velocity dispersion $(\sigma_{vel})$ calculation require a reference point. Instead of using the line's peak, we use the theoretical transition wavelength (for blended/merged lines, the wavelength of the line with the modularity suffix) in the observed frame (accounting for the observation redshift). This approach is advantageous, as exemplified in the complex line fitting shown in Fig. \ref{fig:Halpha_fitting}. In IFS observations with multiple lines, the component with the highest peak may vary significantly as well due to the ionization conditions, making the interpretation of kinematic maps very challenging. Nevertheless, the output logs include both the peak wavelength and the Gaussian centroids of the components. This makes it easy to calculate the radial velocity using other point of reference.

\end{itemize}

%--------------------------------------------------------------------

\section{Discussion}

In the literature, there are outstanding packages avaiilable for the measurement of lines. These include \textsc{IRAF} by \cite{tody_iraf_1986}, \textsc{ALFA} by \cite{wesson_alfa_2016}, \textsc{LZIFU} by \cite{ho_lzifu_2016}, \textsc{LSDCat} by \cite{herenz_lsdcat_2017}, \textsc{GAUSSPY} by \cite{riener_gausspy_2019}, and \textsc{GLEAM} by \cite{stroe_gleam_2021}, among others. Some of these tools, such as \textsc{Pipe3D} by \cite{sanchez_pipe3d_2016} or \textsc{ppxf} by \cite{cappellari_improving_2017}, provide a stellar population synthesis in addition to the line flux measurements. Furthermore, numerous astronomers rely on their custom scripts to quantify spectral features. At this juncture, one might ask how can \lime{} contribute. The authors believe that there are two key tests, which will determine the mid-term success of astronomical software:

    \textbf{Multi-disciplinary design}: The first programming task for many early career astronomers involves accessing and opening digital telescope observations. Afterwards, many will proceed to measure the lines on the spectra. While this is a valuable learning exercise for new astronomers and programmers, this treatment complexity increases very fast, for example, to change the observation type, manage bad pixels or sharing scripts between colleagues. Moreover, this is redundant not only for newcomers but also for their mentors, who are responsible for reviewing the results quality. To avoid these challenges, \lime{} was designed to support both long-slit and IFS data sets, emission and absorption lines and different profile shapes. No matter whether the stellar or galactic study focuses on the chemical or kinematic (or both) analysis, our package can support the procedure.
    
    \textbf{Big data design}: The workload expected from next-generation surveys cannot be managed with the standard algorithms used by the astronomical community. Some of these challenges include remotely processing large IFS data cubes (exceeding 10 GB) or fitting lines with a large number of Gaussian components (more than 15 dimensions). In such scenarios, the data treatment must be simplified to maintain the astronomer's focus on the data interpretation. \lime{} addresses these challenges through standardization. Line labels and bands adhere to a strict format for unique indexing. Fitting properties are defined in external and readable configuration files. Additionally, the multi-level design allows users to set default line types and fittings, with the option to specify specific boundaries for certain objects or lines. This approach significantly reduces the astronomer's coding requirements and facilitates future upgrades, such as implementing machine learning algorithms for line and profile detection. Trained models can confirm line presence and fitting configuration, while these inputs format and workflow remain constant.
     
Finally, \lime{}'s design offers novel options to the community to perform and distribute spectroscopic line measurements.

\subsection{Self-hosted spectroscopic surveys}\label{sec:JWST_virtual}

An observable trend in software development is the migration towards online platforms. This phenomenon, known as Eric Schmidt's Law, is driven by increasing network speeds. In scientific programming applications, this approach offers two significant advantages for the user. First, there is no need to install specialized software, nor its dependencies. Second, complex operations, such as reading data from online platforms, can be centralized and managed by the developer.

The \textsc{Jdaviz} package \cite{developers_jdaviz_2023}, implemented within the Space Telescope Science Institute (STScI) data analysis tools ecosystem, exemplifies this principle. This is a line measuring package, based on Jupyter notebooks, which can be run as an offline application or embedded within a website. While \lime{} is compatible with Jupyter notebooks (indeed, the documentation tutorials\footnote{\textsc{Github} tutorials folder \href{https://github.com/Vital-Fernandez/lime/tree/master/examples/tutorials}{tutorials folder}} are compiled from them), its design is geared towards lower-level functions, which are platform-agnostic. This facilitates the integration into other online platforms, both the current ones and those that may become popular in the future. 

\begin{figure}[h]
\centering
\includegraphics[width=1\columnwidth]{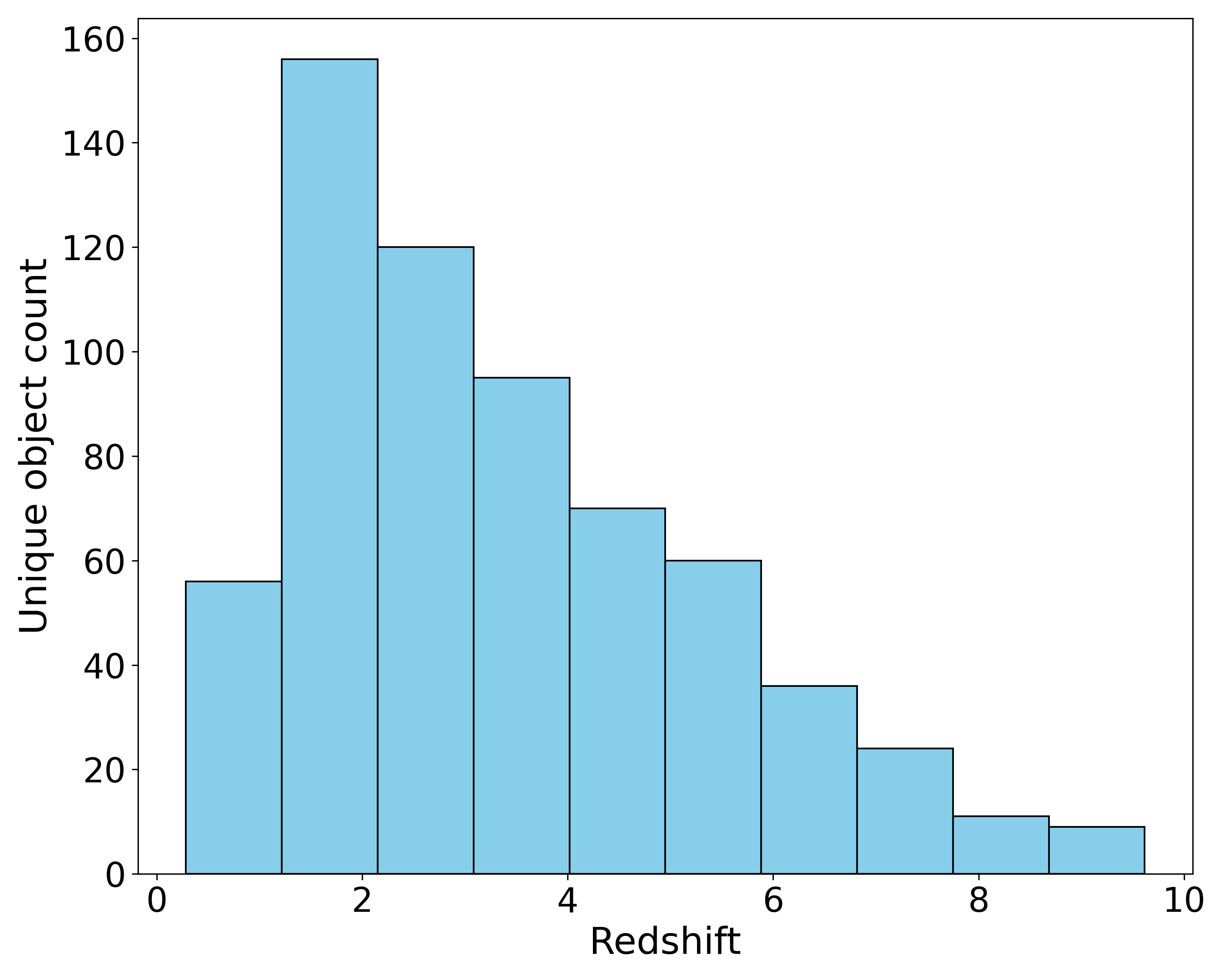}
\caption{\label{fig:histogram_ceers} Histogram showing the number of galaxies in the CEERs virtual observatory with \lime{} flux measurements as a function of redshift.}
\end{figure}

This design is also viable for online platforms, such as the spectroscopy virtual observatory hosted at \href{https://ceers-data.streamlit.app}{ceers-data.streamlit.app}. This site includes spectra observed by the CEERS  survey \citep[see][]{finkelstein_ceers_2022}. This data set corresponds to the \href{https://ceers.github.io/dr07.html}{CEERS v0.7 public release}. CEERS includes multi-object spectroscopic observations of faint galaxies using JWST's Near InfraRed Spectrograph \citep[\textsc{NIRSpec}][]{jakobsen_near-infrared_2022}.  The CEERS NIRSpec observations and data reduction will be described fully in Arrabal Haro et al. (in preparation). This data set includes over 4400 spectra (2D + 1D \emph{.fits} files) for 1114 objects. The CEERS observations used \textsc{NIRSpec}'s medium resolution gratings (R = $\lambda/\Delta\lambda$  $\approx$ 1000,  1.0 < $\lambda(\mu m)$ < 5.1) and its low resolution prism (R $\approx$ 30 to 300) with a wavelength range of 0.6 < $\lambda(\mu m)$ < 5.3.
In this sample, the author has manually identified lines and computed the redshift for 647 unique objects. The histogram in Fig.\ref{fig:histogram_ceers} displays the current number of objects with line measurements as the redshift increases.

This virtual observatory is powered by the \textsc{Streamlit} platform\footnote{CEERs virtual observatory website \href{https://ceers-data.streamlit.app/}{https://ceers-data.streamlit.app/}}. Upon linking a \textsc{streamlit} account with a \textsc{GitHub} repository using this platform API, the website is automatically compiled. The complete code structure is available at the first author's \textsc{Website GitHub}\footnote{CEERs virtual observatory \textsc{GitHub} \href{https://github.com/Vital-Fernandez/ceers-data}{github.com/Vital-Fernandez/ceers-data}}. In this setup, \lime{} is included as a dependency and the widget-based interface is used to pass user inputs to its functions. This enables the user to perform various tasks such as constraining sample selection, plotting spectra, computing redshifts, or inspecting line fittings. For this specific case, the spectra are hosted directly on \textsc{GitHub}. The \href{https://lime-stable.readthedocs.io/en/latest/introduction/api.html#lime.Sample}{Sample} class reads the data directly from the hosting address quickly and with low memory requirements. For situations requiring privacy, the website can be password-protected and/or the \emph{.fits} data encrypted.

These measurements have been used in the works by \cite{arrabal_haro_spectroscopic_2023}, \cite{arrabal_haro_confirmation_2023},  \cite{seille_probing_2023}, \cite{davis_census_2023}, a. At this time, NIRSpec instrument calibration from the Space Telescope
Science Institute (STScI) continues to be updated, and the CEERS team is continuing to improve the fidelity of its data reduction.  There are several known issues with flux calibration, path-loss correction, background subtraction for extended sources, and in some cases with wavelength calibration.  Once the CEERS collaboration is more confident about these calibration issues, it will be possible to download the line measurements from the LiMe website.  Redshift measurements and quality flags based on evaluation by multiple CEERS team members will be published in Arrabal Haro et al. (in preparation).

\subsection{Accuracy and precision} \label{sec:Accuracy_precisison}

An absorption or emission line flux is determined by the number of missing or excess photons from a specific transition. This is quantified by the area covered by the line. As detailed in section \ref{sec:measurements}, there are two types of fluxes per line: one based on Monte Carlo integration and the second on the Gaussian area as given by \ref{eq:Gauss}. Assuming a single line with a Gaussian shape and a linear continuum, four parameters contribute to the line flux measurement and its accuracy: the line amplitude $(A_{gas})$, the noise standard deviation $(\sigma_{noise})$, the gas dispersion velocity $(\sigma_{gas})$, and the spectrum resolution $(\Delta\lambda)$. These parameters can be effectively represented by the dimensionless ratios $A_{gas}/\sigma_{noise}$ and $\sigma_{gas}/\Delta\lambda$. The first ratio traces the S/N of the line, while the second is a measured of the number of pixels covered by the line. Generating and measuring synthetic lines based on these ratios allows us to quantify \lime{}'s (or any algorithm's) accuracy and precision across the entire parameter space. Here, the "true flux" is given by Eq. \ref{eq:Gauss}, while the "true error" can be calculated as:
\begin{equation}
\sigma_{true\,flux}=\sqrt{N_{pix}\cdot\left(\sigma_{noise}\Delta\lambda\right)^{2}}
.\end{equation}
Where $N_{pix}$ is the number of pixels covered by the line. Assuming the Gaussian reaches $4\cdot\sigma_{gas}$ for a given spectrum resolution, the number of pixels is $N_{pix}=2\cdot4\sigma_{gas}/\Delta\lambda$. Rewriting the equation as:
\begin{equation}
\sigma_{true,,flux}=2\sigma_{noise}\sqrt{2\cdot\sigma_{gas}\cdot\Delta\lambda} \label{eq:true_err}
.\end{equation}

\begin{figure*}
\centering
\includegraphics[width=0.495\textwidth]{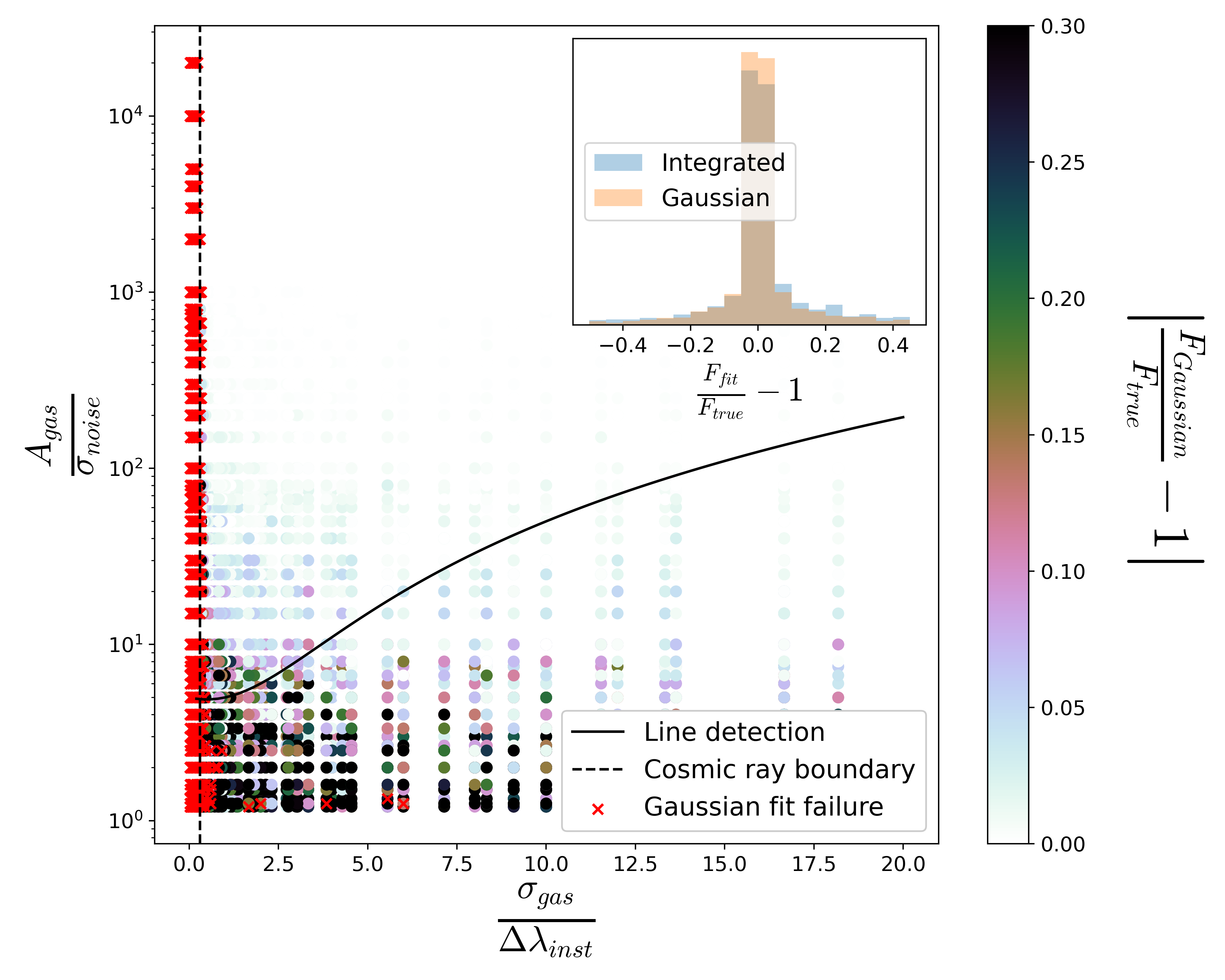}
\includegraphics[width=0.495\textwidth]{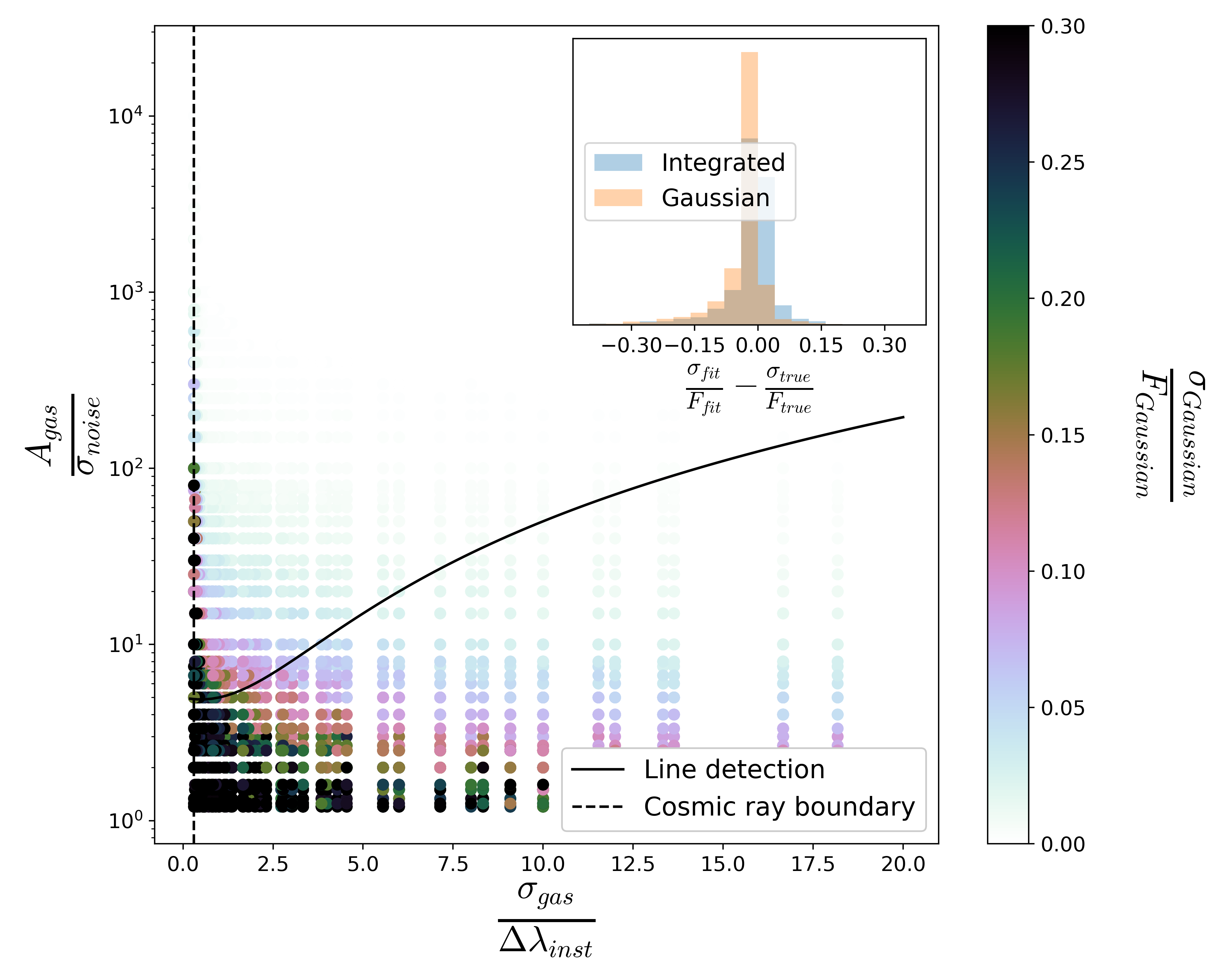}
\caption{\label{fig:accu_pres_gauss} Accuracy and precision evaluation plots for parameters contributing to a line flux's nominal and standard deviation values: line amplitude $(A_{gas})$, standard deviation $(\sigma_{gas})$, line flux noise $(\sigma_{noise})$, and spectrum resolution $(\Delta \lambda_{inst})$. The black solid line indicates the emission line detection boundary, and dashed lines represent the cosmic rays (or dead pixel) "lines" limit. Each circle represents the Gaussian fitting of a synthetic line. Left: Points are color-coded by the absolute relative error between measured and true flux. Inset histogram shows relative error of integrated and Gaussian fluxes. Red crosses indicate points where \lime{} Gaussian fitting failed. Right: Points color-coded by coefficient of variation for Gaussian uncertainty and nominal flux. Inset histogram shows difference between Gaussian and integrated fluxes against true coefficient of variation.}
\end{figure*}

Fig. \ref{fig:accu_pres_gauss} shows the relative error and coefficient of variation for emission line fitting tests. The black solid line marks the visual detection boundary, detailed in Section \ref{sec:line_detection}. Scatter points represent the parameter space coordinates for generated and measured emission lines, with color-coding indicating the magnitude of absolute relative error for measured Gaussian fluxes and the coefficient of variation for Gaussian uncertainty. Red crosses mark points where \lime{} Gaussian fitting failed, mostly below the "cosmic ray boundary" where lines occupy a single pixel. These failed measurements, where the minimizer does not converged, have the profile measurements (see Fig. \ref{tab:measurements}) output as \lstinline{nan}.  By default \lime{} settings impose the spectrum pixel width as the lower limit for $\sigma_{gas}$, leading to automatic failure in these cases. However, in some instruments, lines might have this cosmic ray profile. Therefore, it may be necessary to ease this constraint. Additional failures are observed well below the line detection boundary, likely due to noisy spectra complicating Gaussian fitting.

The plots show in general a very good accuracy and precision (less than 5\%) above the line detection boundary. The most uncertain region is where $A_{gas}/\sigma_{noise} < 10$ and $\sigma_{gas}/\Delta\lambda < 2$, corresponding to lines close to noise level and narrower than 12 pixels. Here, measurement accuracy is dominated by random error, with relative uncertainty up to 30\%. However, this is not unique to \lime{} measurements, as emission lines with $S/N<5$ (see Eq. \ref{eq:SN_rola}) are subject to a strong positive bias \citep[see][]{rola_estimation_1994}. This means that in most cases the line intensity measured is higher than the true one. Assuming $A_{gas}/\sigma_{noise} \approx S/N$, if we take the points in Fig.\ref{fig:accu_pres_gauss} with $A_{gas}/\sigma_{noise} \leq 5$, but above the detection boundary (Eq. \ref{eq:empiric_line_detection}) we observe a skewed distribution for the Gaussian fluxes peaking around 10\% higher than the true values. In contrast, the integrated fluxes seem to be closer to the true values but this distribution also has large wings. In our tests, where an emission line always exists, the relative error stabilizes below 5\% when $A_{gas}/\sigma_{noise} \gtrsim 13$. The same pattern is observed with Monte Carlo measurements, as shown in the left inset histogram, comparing density distributions for two flux measurements. While integrated fluxes might be slightly larger than Gaussian for some regions, this is influenced by line band selection. On the right diagram of Fig. \ref{fig:accu_pres_gauss}, the plot structure is similar but color-coded by coefficient of variation. Here, Gaussian fittings tend to report slightly smaller uncertainties than expected, indicating more stable measurements for weaker lines but with a tendency to underestimate uncertainties.

\subsection{Line detection and measurement issues} \label{sec:line_detection}

The first published work using \lime{} was the chemical analysis of the galaxy CGCG007-025 in \cite{fernandez_resolved_2023}. Since the beta release in that manuscript, feedback from collaborators has highlighted two key areas for \lime{}'s future development. The first issue concerns the fitting of blended lines with four or more transitions. For example, a multi-Gaussian profile fitting is a non-linear process, and as such, exploring the parameter space for an optimal solution becomes increasingly complex with higher dimensions. Fittings, such as those shown in Fig. \ref{fig:Halpha_fitting}, are very sensitive to initial conditions and may not converge at all. Packages such as \textsc{ppxf} or \textsc{Gandalf} linearize this model by simultaneously fitting several transitions with a unique radial velocity in the line of sight and dispersion velocity per kinematic component. Alternatively, \lime{} can export kinematics from one line to another to reduce the number of effective dimensions. However, as demonstrated in \cite{amorin_complex_2012}, high-resolution observations of strong emission lines like $H\alpha$ can have many components, both narrow and broad. This makes the finding the solution very time-consuming. In \cite{fernandez_bayesian_2019}, we introduced a novel methodology using neural networks to fit the complete chemical space parameters under the direct method paradigm. This approach was expanded in \cite{fernandez_new_2021} and \cite{fernandez_resolved_2023} to fit photo-ionization models. Future work will explore the use of neural networks for fitting complex multi-Gaussian profiles.

\begin{figure}
\centering
\includegraphics[width=1\columnwidth]{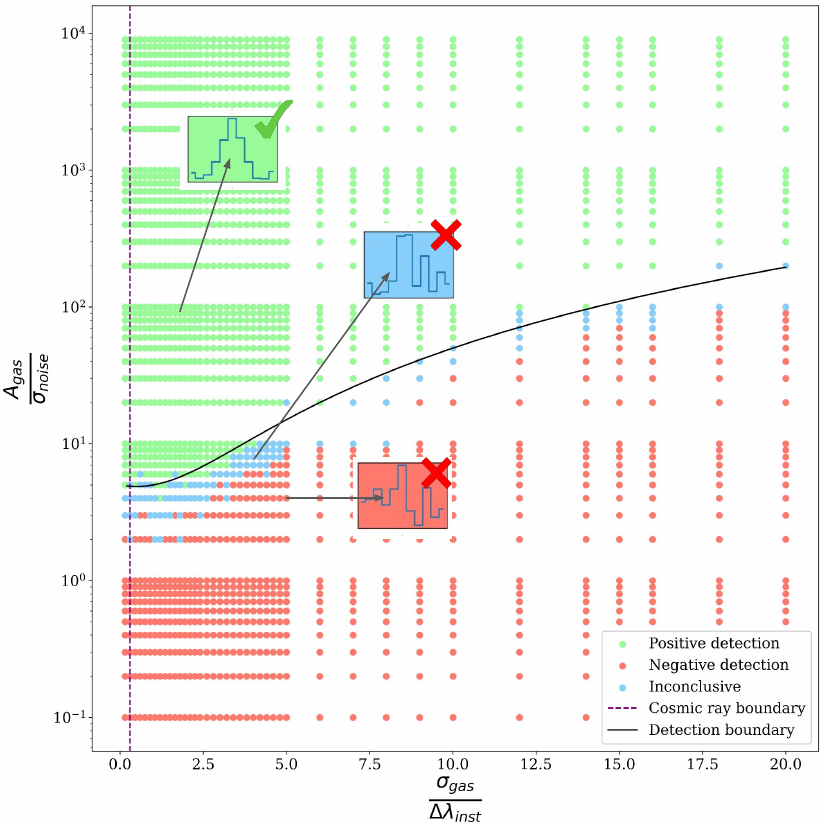}
\caption{\label{fig:parameter_space_line_detection} Parameter space for line measurement as described in the text, with scatter points color-coded according to a visual criterion for line detection. In the color version, these are green for positive detection, red for negative, and blue for inconclusive. The inset plots show examples of each case. The black solid line represents the empirical detection boundary}
\end{figure}

Early feedback also identified \lime{}'s main user error source: attempting to measure a non-existent line or kinematic component. The line detection procedure outlined in Section \ref{sec:line_detec} is effective as long as the user provides a good continuum parametrization and a reasonable list of expected lines. However, it cannot autonomously confirm the presence of lines. Ideally, three independent diagnostics should be compared to confirm a line's presence. In Fig. \ref{fig:parameter_space_line_detection}, we revisit the parameter space from the previous sub-section. For the given parameter range, Gaussian curves are generated and tagged with positive (green), negative (red), or inconclusive (blue) based on the author's visual criteria. This exercise highlights the region where line detection is expected. The boundary can be approximated by:
\begin{equation}
\left(\frac{A_{line}}{\sigma_{noise}}\right)_{detection}=\frac{1}{2}\left(\frac{\sigma_{line}}{\Delta\lambda_{inst}}\right)^{2}-\frac{1}{2}\frac{\sigma_{line}}{\Delta\lambda_{inst}}+5\label{eq:empiric_line_detection}
.\end{equation}

This empirical model can label synthetic data for training machine learning algorithms: users provide a range of parameters to generate lines (detailed as in Sect. \ref{sec:Accuracy_precisison}), then use Eq. \ref{eq:empiric_line_detection} to label data arrays with positive line detections. A supervised learning algorithm \citep[see][and references therein]{baron_machine_2019} can then be used to create a model for line detection in spectral bands. Initial tests with binary classifiers have been successful in predicting lines. The main advantage from this approach is that it requires very little input from the user for an efficient line detection. However, unlike the traditional technique described in Sect. \ref{sec:line_detection}, trained-models can be very sensitive to the expected. For example, a model trained with Gaussian lines, will fail to detect blended lines even if their signal to noise is very high. Currently, the authors are exploring a chain of machine-learning models to reproduce the human vision pattern (seen in Fig. \ref{fig:parameter_space_line_detection}), while avoiding these pitfalls.

\subsection{Benchmarks}

\begin{figure}
\centering
\includegraphics[width=1\columnwidth]{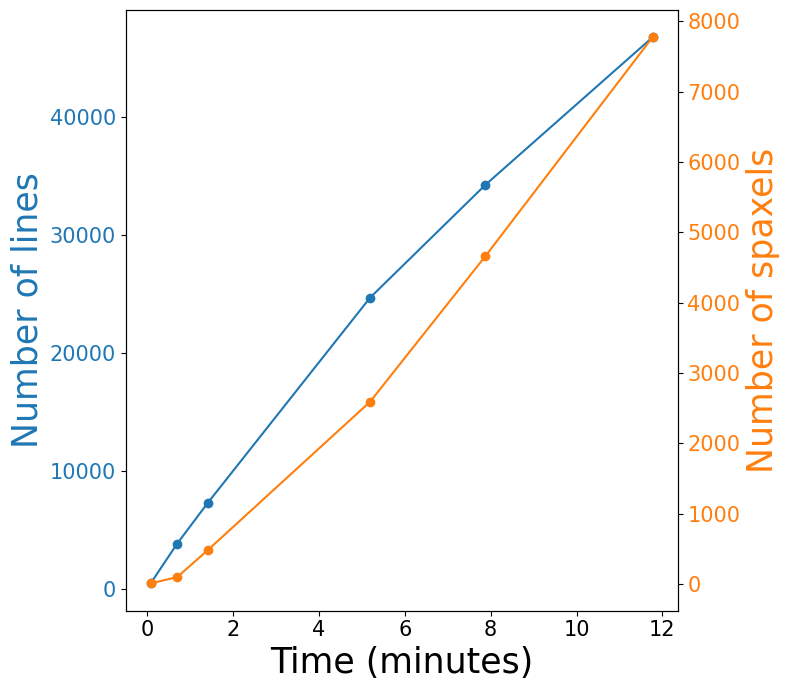}
\caption{\label{fig:benchmarks} \lime{} benchmarks for the analysis of CGCG007-025 described in \protect\cite{fernandez_resolved_2023} with the cumulative sum of lines and spaxels. The 6 masks have [459, 3418, 3431, 17390, 9579, 12498] spaxels. At each mask, we measure [96, 91, 81, 77, 59, 54] lines per second for [2,  2,  9,  9, 13, 13] spaxels per second.}
\end{figure}

To assess the performance of \lime{}, a script is provided on the library's \textsc{GitHub}\footnote{\href{https://github.com/Vital-Fernandez/lime/tree/master/examples/benchmark}{github.com/Vital-Fernandez/lime/examples/benchmark}} for fitting the MUSE observation of the galaxy CGCG007-025. This is a low-mass blue compact dwarf \citep[$log(M_{*})=8.17$, see][]{gavilan_chemical_2013}, undergoing a galaxy-wide star-forming burst. In \cite{fernandez_resolved_2023}, we analyzed its chemical composition, focusing on 6 spatial regions encompassing 7774 spaxels. These regions display an ionization gradient, from the hard radiation at the central stellar cluster, to the low ionization regions in the galaxy outskirts, where only $H\alpha$, $H\beta$, and the $[OIII]4959,5007\AA$ doublet are observed. Unlike the emission line quantity, the number of spaxels per mask increases, as we move to lower ionization regions. Detailed fitting properties are described in that manuscript, but the measurement performance is illustrated in Fig.\ref{fig:benchmarks}. The left (blue) ordinate axis shows the cumulative number of lines, while the right (orange) one corresponds to the number of spaxel spectra. The complete cube analysis took 11.77 minutes for 46775 line measurements. Fig.\ref{fig:benchmarks} details the rate of lines and spaxels measured per second. These rates vary because the computational load is split between two tasks: line tasks (profile fitting and parameters calculation) and spaxel tasks (spectrum extraction, line detection, data formatting, and writing to memory). Fewer lines mean that spaxel tasks dominate the computational time. These tests were performed on a mid-to-high range desktop processor Intel Core i5-13600KF 3.5 GHz 14-Core.

To gauge \lime{}'s efficiency, we can compare it with a state-of-the-art library like \textsc{ALFA} by \cite{wesson_alfa_2016}. 
This \textsc{Fortran} based package was developed to analyze emission-rich, high-resolution spectra of planetary nebulae. The library's paper includes benchmarks for a 5.1 Gb MUSE cube, where 41022 spaxels were treated, resulting in over 2 million emission line measurements in 20 hours. This translates to about 49 lines per spaxel, similar to our first mask in Fig.\ref{fig:benchmarks}, suggesting \lime{} would take approximately 5.7 hours for a comparable observation. While this is a simplistic comparison involving different observations, CPUs, line profile types, and likely, file-saving procedures, it's reasonable to conclude that \lime{} is on par, performance-wise, with highly optimized packages.

Beyond quantitative benchmarks, a more realistic approach consists in setting efficiency targets. There are three scenarios, where astronomers have to deal with Big Data. The first one involves individual researchers. For a scientific manuscript discussing the results from tens of thousand of spectra (as in Fig.\ref{fig:benchmarks}) a few hours for the line measurements are negligible. The second scenario involves the collaboration between several researchers across institutions. If we scale the previous data size by the number of researchers, even a few days of measurements are acceptable for the workflow. In this scenario, however, the main challenges involve distributing the data among the collaborators so it can be reviewed. As we show in Fig. \ref{fig:benchmarks} and describe in Sect. \ref{sec:JWST_virtual}, the \lime{} design is particularly well positioned to address these two scenarios. The third case is the analysis of large surveys. At this point, the software needs to treat the data as fast as it is produced (observation + calibration). In this case, \lime{} will need to explore performance-enhancing techniques such as CPU/GPU parallelisation or machine learning based models. Before this is possible, however, it is necessary to enhance the data workflow to make sure it can handle very large data files. Currently, Pignata et al (in preparation), are using \lime{} to measure very large \textsc{MUSE} data sets (> 40 GBs) within a CPU cluster. From this feedback, we will explore advanced techniques to handle the data reading, management and writing.

%--------------------------------------------------------------------

\section{Conclusions}

In this manuscript, we introduce a new LIne MEasuring package: \lime{}. This library boasts a simple yet flexible design, accommodating both long-slit and integral field spectroscopy (IFS) data. It is freely available, comprehensively documented with tutorials, measurements description, and an API detailing the commands and their inputs. Additionally, \lime{} features commands to plot and interact with the measurements. The library easy installation, modular design, and competitive performance further enhance its usability. We also discuss various topics to engage the community in addressing Big Data challenges in astronomy and to foster \lime{} future development:

\begin{itemize}
    \item \lime{} is designed to support both chemical and kinematic studies. We present a notation system for describing transitions, which is both human and machine readable and provides a flexible format for specifying transition particles, wavelengths, kinematic components, or profile types. For multiple species transitions or kinematic components, this notation distinguishes between lines that can be isolated mathematically (blended) and those that cannot (merged). \lime{} relies on a set of bands to locate lines and derive flux uncertainty. We include a line bands database, which the user can adjust to their observations

    \item The profile fittings in \lime{} support a wide range of boundaries. Users can fix parameter values, establish limits, and define relationships between them, including inequality relations. The transition components and fitting boundaries can be specified in an external, human-readable file, eliminating coding requirements. For handling multiple spectra or IFS data cubes, a multi-level fitting configuration simplifies user input by establishing global and local fitting configurations.

    \item \lime{}'s modern implementation makes it suitable for state-of-the-art interactive and programmatic platforms. As a practical example, we showcase a spectroscopic virtual observatory at \href{https://ceers-data.streamlit.app}{ceers-data.streamlit.app}. This repository contains JWST NIRSpec observations from the CEERs survey for a total of 4408, 1D and 2D sepctra. Using \lime{}, we have measured redshift in 647 unique objects within a range of $0.29 \leq z \leq 9.62$. While direct data download is not currently available, users can inspect individual fittings, illustrating how \lime{} can be used to facilitate the access to spectroscopic data.

    \item In our dicussion of \lime{}'s accuracy and precision, we propose a parameter space normalization that highlights the specific S/N and instrument resolution regions where line measurements can be challenging. In regimes where $A_{gas}/\sigma_{noise} \lesssim 13$ and lines consist of fewer than 12 pixels, users should be wary of random errors dominating the flux. At this region, Gaussian fluxes provide more accurate measurements, while integrated errors better approximate the intrinsic uncertainty.  

    \item Building on this normalization, we suggest an empirical relation to define the $A_{gas}/\sigma_{noise}$ detection limit as a function of $\sigma_{gas}/\Delta\lambda_{inst}$. This relation can label synthetic lines in samples to train supervised machine learning models. We are currently exploring the best approach to integrate this methodology into \lime{}, with an aim to reduce the workload for astronomers.

\end{itemize}

Future \lime{} development will take into consideration the requests from the community to improve its workflow and add new features.

\begin{acknowledgements}
      The authors want to thank the anonymous referee for his/her comments. V.F. acknowledges financial support provided by FONDECYT grant 3200473. V.F. also acknowledges the support by the Eric and Wendy Schmidt AI in Science Postdoctoral Fellowship, a Schmidt Futures program, at the Michigan Institute for Data Science, University of Michigan. R.A. acknowledges the support of ANID FONDECYT Regular Grant 1202007 and DIDULS/ULS PTE2053851. C.M. acknowledges the support of UNAM/DGAPA/PAPIIT grant IG101223. V.F. wants to thank all the early \lime{} users for their feedback. Finally, V.F. wants to thank Juan Luis Cano Rodríguez for his support and insight in \lime{} documentation.
      
\end{acknowledgements}

\bibliographystyle{aa}
\bibliography{references.bib}


\end{document}